%% file: RR-6726.tex
\documentclass[a4paper]{article}
\usepackage{RR}

\RRdate{\today}

\RRauthor{Pierre Genev\`es\thanks{CNRS}
   \and
Nabil Laya\"ida
}

\authorhead{Genev\`es, Laya\"ida, \& Quint}

\RRtitle{Analyseur Statique pour XML et XPath: Manuel Utilisateur}
\RRetitle{XML Static Analyzer User Manual}
\titlehead{XML Reasoning Solver User Manual}

\RRresume{Ce manuel documente l'utilisation du solveur logique d\'ecrit dans\cite{geneves-phd06,geneves-pldi07}.}
\RRabstract{This manual provides documentation for using the logical solver introduced in \cite{geneves-phd06,geneves-pldi07}.}
\RRmotcle{Analyse Statique, Logique, Solveur,  Satisfaisabilit\'e, XML, Schema, Requ\^etes, XPath}
\RRkeyword{Static Analysis, Logic, Satisfiability Solver, XML, Schema, XPath, Queries}
 
\RRprojets{Wam}
\RRtheme{\THSym} 

\RCGrenoble

\usepackage{named}
\usepackage{amsmath,amssymb}
\usepackage{stmaryrd} 
\usepackage{tikz} 
  
\usepackage{pgf} 
\usepackage{multirow}
\usepackage{fancyvrb}

\input{Markup}

\begin{document}   
\RRNo{6726} 
\makeRR 

%

%






%
%

%


%

\section{Introduction}
This document describes the logical solver introduced in \cite{geneves-phd06,geneves-pldi07} and provides informal documentation for using its implementation. 

The solver allows automated verification of properties that are expressed as logical formulas over trees.  A logical formula may for instance express structural constraints or navigation properties (like e.g. path existence and node selection) in finite trees.

A decision procedure for a logic traditionally defines a partition of the set of logical formulas: formulas which are \emph{satisfiable} (there is a tree which satisfies the formula) and remaining formulas which are \emph{unsatisfiable} (no tree satisfies the given formula). Alternatively (and equivalently), formulas can be divided into \emph{valid} formulas (formulas which are satisfied by all trees) and \emph{invalid} formulas (formulas that are not satisfied by at least one tree). 
The solver is a satisfiability-testing solver: it allows checking satisfiability (or unsatisfiability) of a given logical formula. Note that validity of a formula $\phi$ can be checked by testing $\neg \phi$ for unsatisfiability.

The solver can be used for reasoning over finite ordered trees whatever these trees do actually represent. In particular, the logic and the solver are specifically adapted for formulating and solving problems over XML tree structures  \cite{xml}.  The logic can express navigational properties like those expressed with the XPath  standard language \cite{xpath} for navigating and selecting sets of nodes from XML trees. Additionally,  the logic is expressive enough to encode any regular tree language property (it subsumes finite tree automata). It can encode constraints definable with common XML tree type definition languages (such as DTD \cite{xml}, XML Schema \cite{xml-schemas}, and Relax NG \cite{relax}).  The logic provides high-level constructs specifically designed for reasoning directly with such XML concepts: the user can directly write an expression using XPath notation in the logic, or even refer to an XML type in the logic. These characteristics make the system especially useful for solving problems like those encountered in the static analysis of XML code, static verification of XML access control policies,  XML data security checking, XML query optimization,  and the construction of static type-checkers, and optimizing compilers for a wide variety of tree-manipulating programs and XML processors.


\paragraph{Outline}
This user manual is organized as follows: Section~\ref{getting-started} describes the basics for using the solver without requiring any logical knowledge; Section~\ref{logical-insights} gives some insights on the logic, especially on the simple yet general data tree model used by the logic (Section~\ref{data-model}) and on the syntax of logical formulas (Section~\ref{formulas-syntax}) including high-level constructs for embedding XPath expressions and XML tree types directly in the logic. Finally, Section~\ref{theory-overview} provides an overview of the background theory underlying the logic and its solver, with related references.

\section{Getting Started with XML Applications} \label{getting-started}

The logical solver is shipped as a compressed file which, once extracted, provides binaries along with all required libraries. The  ``\texttt{solver.jar}'' executable file takes a filename as a parameter\footnote{Running the command ``\texttt{java -jar solver.jar}'' prints the list of required and optional arguments.}. The filename refers to a text file containing the logical formula to solve. 
For example, provided a recent\footnote{A Java virtual machine version 1.5.0 (or further) and a Java compiler compliance level version 5.0 (or further).} Java runtime engine is installed, the following command line: 

\begin{Verbatim}[frame=single, fontsize=\small]
java -jar solver.jar formula.txt
\end{Verbatim}

\noindent runs the solver on the logical formula contained in ``\texttt{formula.txt}''. The full syntax of logical formulas is given in Section~\ref{formulas-syntax}. The following examples introduce the logical formulation of some simple yet fundamental XML problems, and how the solver output should be interpreted.

\paragraph{Example 1:  emptiness test for an XPath expression.}
The most basic decision problem for a query language is the emptiness test of an expression: whether or not a query is self contradictory and always yields an empty result. This test is important for error-detection and optimization of host languages implementations, \ie implementations that process languages in which XPath expressions are used. For instance, if one can decide at compile time that a query result is empty then subsequent bound computations can be ignored. For checking emptiness of the XPath expression \texttt{a/b[following-sibling::c/parent::d]}, the contents of the ``\texttt{example1.txt}'' file simply consists of the following line: 

\begin{Verbatim}[frame=single, fontsize=\small, label={\texttt{example1.txt}}]
select("a/b[following-sibling::c/parent::d]")
\end{Verbatim}

\noindent Running the solver with `` \texttt{example1.txt}'' as parameter yields the following trace:

\begin{Verbatim}[frame=single, fontsize=\small, label={Output for \texttt{example1.txt}}]
Reading example1.txt

Satisfiability Tested Formula:
(mu X5.(((b & (mu X2.(<-1>(a & (mu X1.(<-1>T | <-2>X1))) | <-2>X2))) 
& (mu X4.(<2>((mu X3.(<-1>d | <-2>X3)) & c) | <2>X4)))|(<1>X5|<2>X5)))

Computing Relevant Closure
Computed Relevant Closure [1 ms].
Computed Lean [1 ms].
Lean size is 20. It contains 14 eventualities and 6 symbols.
Computing Fixpoint.....[4 ms].
Formula is unsatisfiable [14 ms].
\end{Verbatim}


\noindent The input XPath expression is first parsed and compiled into the logic.  The corresponding logical translation whose satisfiability is going to be tested is printed. The solver then computes the Fisher-Ladner closure and the Lean of the formula: the set of all basic subformulas that notably defines the search space that is going to be explored by the solver (see \cite{geneves-pldi07} for details). The solver attempts to build a satisfying tree in a bottom-up way, in the manner of a fixpoint computation that iteratively updates a set of tree nodes. This computation is performed in at most $2^{O(n)}$ steps with respect to size $n$ of the Lean. 

In this example, no satisfying tree is found: the formula is unsatisfiable (in other terms, no matter on which XML document this XPath expression is evaluated, it will always yield an empty result). Intuitively, that is because this XPath expression contains a contradiction: according to the query, the same node is required to be named both ``\texttt{a}'' and ``\texttt{d}'', which is not allowed for an XML tree. 

Empty queries often come from the use of an XPath expression in a constrained setting. The combination of navigational information of the query and structural constraints imposed by a DTD (or XML Schema) may rapidly yield contradictions.  Such contradictions can also be detected by checking a logical formula for satisfiability.

\paragraph{Example 2: checking XPath emptiness in the presence of tree constraints.} \label{example-xpath-sat}
Suppose we want to check emptiness of the XPath expression

\begin{Verbatim}[frame=single, fontsize=\small]
descendant::switch[ancestor::head]/descendant::seq/
   descendant::audio[preceding-sibling::video]
\end{Verbatim}

\noindent over the set of documents defined by the DTD of the SMIL language \cite{smil}. The following formula is used:

\begin{Verbatim}[frame=single, fontsize=\small, label={\texttt{example2.txt}}]
select("descendant::switch[ancestor::head]/descendant::seq/
   descendant::audio[preceding-sibling::video]",
      type("sampleDTDs/smil.dtd", "smil"))
\end{Verbatim}

\noindent The first argument for the predicate \texttt{type()} is a path to the DTD file
(here the DTD is assumed to be located in a subdirectory called ``sampleDTDs''), and the second argument is the name of the element to be considered as top-level start symbol. Running the solver with this ``\texttt{example2.txt}'' file as parameter yields the following trace:

\begin{Verbatim}[frame=single, fontsize=\small,  label={Output for \texttt{example2.txt}}]
Reading example2.txt
Converted tree grammar into BTT [169 ms].
Translated BTT into Tree Logic [60 ms].

Satisfiability Tested Formula: 
(mu X22.(((audio & (mu X20.(<-1>((seq & (mu X19.(<-1>(((switch
 & (mu X17.(<-1>(
(let_mu
  X1=(((meta & ~(<1>T)) & ~(<2>T)) | ((meta & ~(<1>T)) & <2>X1)),
  ...
  X16=((smil & (~(<1>T) | <1>X15)) & ~(<2>T))
in
  X16) | X17) | <-2>X17))) & (mu X18.(<-1>(head | X18) | <-2>X18))) 
  | X19) | <-2>X19)))  | X20) | <-2>X20))) & 
  (mu X21.(<-2>video | <-2>X21)))  | (<1>X22 | <2>X22)))

Computing Relevant Closure
Computed Relevant Closure [39 ms].
Computed Lean [1 ms].
Lean size is 50. It contains 31 eventualities and 19 symbols.
Computing Fixpoint......[37 ms].
Formula is satisfiable [99 ms].
A satisfying finite binary tree model was found [52 ms]:
smil(head(switch(seq(video(#, audio), layout), meta), #), #)
In XML syntax:
<smil xmlns:solver="http://wam.inrialpes.fr/xml" solver:context="true">
  <head>
    <switch>
      <seq>
        <video/>
        <audio solver:target="true"/>
      </seq>
      <layout/>
    </switch>
    <meta/>
  </head>
</smil>
\end{Verbatim}

\noindent The referred external DTD (tree grammar) is first parsed, converted into an internal representation on binary trees (called ``BTT'' and that corresponds to the mapping described in~\ref{data-model}), and then compiled into the logic. The XPath expression is also parsed and compiled into the logic so that the global formula can be composed. In that case, the formula is satisfiable (the XPath expression is non-empty in the presence of this DTD). The solver outputs a sample tree for which the formulas is satisfied. This sample tree is enriched with specific attributes:  the ``solver:target'' attribute marks a sample node selected by the XPath expression when evaluated from a node marked with ``solver:context''.

\paragraph{Example 3: checking containment and equivalence between XPath expressions.}
One of the most essential problem for a query language is the containment problem: whether or not the result of one query is always included into the result of another one. Containment for XPath expressions is for instance needed for the static type-checking of XPath queries, for the control-flow analysis of XSLT \cite{xslt}, for checking integrity constraints in XML databases, for XML data security... 

\noindent Suppose for instance that we want to check containment between the following XPath expressions:
\begin{Verbatim}[frame=single, fontsize=\small]
descendant::d[parent::b]/following-sibling::a
\end{Verbatim}
\noindent and:
 \begin{Verbatim}[frame=single, fontsize=\small]
 ancestor-or-self::*/descendant-or-self::b/a[preceding-sibling::d]
 \end{Verbatim}
 \noindent Since containment corresponds to logical implication,  we actually want to check whether the implication of the two corresponding formulas is valid.  Since we use a satisfiability-testing algorithm, we verify this validity by checking for the unsatisfiability of the negated implication, as follows:
\begin{Verbatim}[frame=single, fontsize=\small, label={\texttt{example3.txt}}]
~( select("descendant::d[parent::b]/following-sibling::a",#)
  => select("ancestor-or-self::*/descendant-or-self::b
                                    /a[preceding-sibling::d]",#))
\end{Verbatim}

\noindent Note that XPath expressions must be compared from the same evaluation context, which can be any set of nodes, but should be the same set of nodes for both expressions. This is denoted by ``\texttt{\#}''. Running the solver with this ``\texttt{example3.txt}''  file results in the following trace:

\begin{Verbatim}[frame=single, fontsize=\small, label={Output for \texttt{example3.txt}}]
Reading example3.txt

Satisfiability Tested Formula:
(mu X26.(((a & (mu X15.((<-2>T & (~(<-2>T) | <-2>((d & (mu X13.((<-1>T
& (~(<-1>T) | <-1>(_context | X13))) | (<-2>T & (~(<-2>T) | <-2>X13)))))
& (mu X14.((<-1>T & (~(<-1>T) | <-1>b)) | (<-2>T & (~(<-2>T)
 | <-2>X14))))))) | (<-2>T & (~(<-2>T) | <-2>X15))))) & ((~(a) | 
 (mu X22.((~(<-1>T) | <-1>(~(b) | ((~(_context) & (~(<1>T) | 
 <1>(mu X18.((~(_context) & (~(<1>T) | <1>X18)) & (~(<2>T) | 
 <2>X18))))) & (mu X20.((~(<-1>T) | <-1>((~(_context) & (~(<1>T) | 
 <1>(mu X19.((~(_context) & (~(<1>T) | <1>X19)) & (~(<2>T) | 
 <2>X19))))) & X20)) & (~(<-2>T) | <-2>X20)))))) &(~(<-2>T) | 
 <-2>X22)))) | (mu X25.((~(<-2>T) | <-2>~(d)) & (~(<-2>T) |
 <-2>X25))))) | (<1>X26 | <2>X26)))

Computing Relevant Closure
Computed Relevant Closure [4 ms].
Computed Lean [1 ms].
Lean size is 29. It contains 23 eventualities and 6 symbols.
Computing Fixpoint.....[8 ms].
Formula is unsatisfiable [22 ms].
\end{Verbatim}

\noindent The tested formula is unsatisfiable (in other terms: the implication is valid), so one can conclude that the first XPath expression is contained in the second XPath expression.

A related decision problem is the equivalence problem: whether or not two queries always return the same result. It is important for reformulation and optimization of an expression, which aims at enforcing operational properties while preserving semantic equivalence.  Equivalence is reducible to containment (bi-implication) and is noted \texttt{<=>} in the logic. Note that the previous XPath expressions are not equivalent. The reader may check this by using the solver, that will generate the following counter-example tree:

\begin{center}
\begin{small}
\begin{Verbatim}[frame=single]
<b xmlns:solver="http://wam.inrialpes.fr/xml">
  <d/>
  <a solver:context="true" solver:target="true"/>
</b>
\end{Verbatim}
\end{small}
\end{center}

\section{Logical Insights}
\label{logical-insights}

\subsection{Data Model for the Logic}
\label{data-model}
An XML document is considered as a finite tree of unbounded depth and arity, with two kinds of nodes respectively named elements and attributes.  In such a tree, an element may have any number of children elements, and may carry zero, one or more attributes. Attributes are leaves. Elements are ordered whereas attributes are not, as illustrated on Figure~\ref{fig:xml-tree}. The logic allows reasoning on such trees. Notice that from an XML perspective, data values are ignored. 

\definecolor{mydarkblue}{rgb}{0,0.08,0.45} 
\newcommand{\changingcolor}{gray}
\newcommand{\metiquette}[1]{\begin{tiny}{#1}\end{tiny}}
\newcommand{\setiquette}[1]{\begin{small}{#1}\end{small}}
\newcommand{\labelstyle}[1]{\texttt{#1}}
\newcommand{\attdefvalue}{\textvisiblespace}
\newcommand{\attcolor}{mydarkblue}
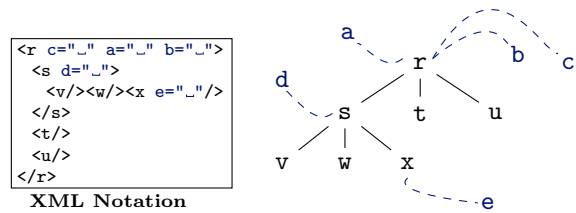
\begin{figure}[h]
\begin{center}
\begin{tikzpicture}[scale=1.1] at (0,0)

\begin{scope}[xshift=-3cm, yshift=-0.9cm, scale=0.6]
 \draw [black, font=\scriptsize, text badly ragged, at start, text width=5cm]  (16,4) node {\texttt{<r \textcolor{\attcolor}{c="\attdefvalue" a="\attdefvalue" b="\attdefvalue"}>} \\~~\texttt{<s \textcolor{\attcolor}{d="\attdefvalue"}>} \\~~~~\texttt{<v/><w/><x \textcolor{\attcolor}{e="\attdefvalue"}/>} \\~~\texttt{</s>} \\~~\texttt{<t/>}\\~~\texttt{<u/>} \\\texttt{</r>}};
 \draw [black] (12.1,5.5) -- (16.5,5.5) -- (16.5,2.5) -- (12.1,2.5) -- cycle;
 \draw[black, font=\scriptsize, text badly ragged, at start, text width=5cm] (16.25, 2.25) node (not) {\textbf{XML Notation}};
\end{scope}
		
\begin{scope}[xshift=0cm, yshift=0cm, scale=0.6]
	\draw [\attcolor, dashed]  (14.9,3.6) .. controls +(40:-0.5cm) and +(40:0.1cm) .. (14,3.9); 
	\draw (13.8, 4.1) node (att3) [text= \attcolor] {\labelstyle{a}};

	\draw [\attcolor, dashed]  (15.5,3.6) .. controls +(40:1.5cm) and +(-40:-0.1cm) .. (17,3.9); 
	\draw (17.2, 3.7) node (att1) [text= \attcolor] {\labelstyle{b}};
	
	\draw [\attcolor, dashed]  (15.5,3.6) .. controls +(60:2.5cm) and +(4:-0.3cm) .. (18,3.7); 
	\draw (18.2, 3.5) node (att2) [text= \attcolor] {\labelstyle{c}};
	
	\draw [\attcolor, dashed]  (13.5,2.5) .. controls +(40:-0.5cm) and +(40:0.1cm) .. (12.5,3); 
	\draw (12.5, 3.2) node (att4) [text= \attcolor] {\labelstyle{d}};

	\draw [\attcolor, dashed]  (15,1.2) .. controls +(40:-0.5cm) and +(-5:0.1cm) .. (16.4,0.7); 
	\draw (16.6, 0.7) node (att4) [text= \attcolor] {\labelstyle{e}};

	\draw (15.25, 3.5) node (r) [text=black] {\labelstyle{r}};

	\draw (13.75, 2.5) node (s1) [text=black] {\labelstyle{s}};
	\draw (15.25, 2.5) node (s2) [text=black] {\labelstyle{t}};		
	\draw (16.75, 2.5) node (s3) [text=black] {\labelstyle{u}};  

	\draw (12.5, 1.5) node (f1) [text=black] {\labelstyle{v}};
	\draw (13.75, 1.5) node (f2) [text=black] {\labelstyle{w}};
	\draw (15, 1.5) node (f3) [text=black] {\labelstyle{x}};
	
	\draw (r) -- (s3);
	\draw (r) -- (s1);
	\draw (r) -- (s2);
	\draw (s1) -- (f3);
	\draw (s1) -- (f2);
	\draw (s1) -- (f1);

\end{scope}
\end{tikzpicture}
\end{center}
\caption{Sample XML Tree with Attributes.}\label{fig:xml-tree}
\end{figure}

\begin{figure}
\includegraphics[keepaspectratio, width=11cm]{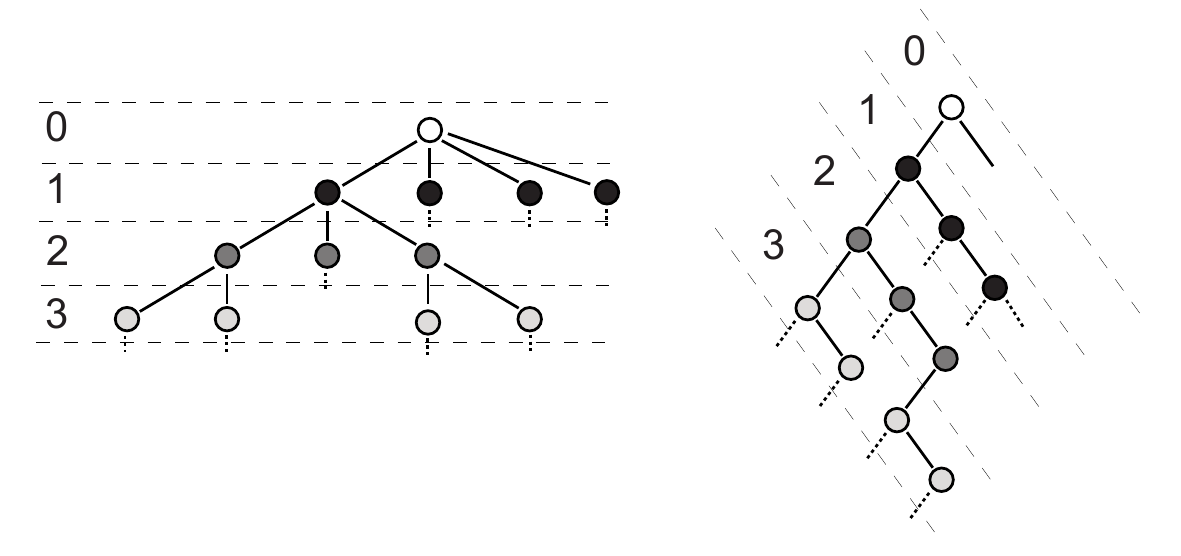}
\caption{Binary Encoding Principle.}\label{fig:binary-encoding}
\end{figure}

\paragraph{Unranked and Binary Trees}
There are bijective encodings between unranked trees (trees of unbounded arity) and binary trees. Owing to these encodings binary trees may be used instead of unranked trees without loss of generality. The logic operates on binary trees.
The logic relies on the ``first-child \& next-sibling'' encoding of unranked trees. In this encoding, the first child of a node is preserved in the binary tree representation, whereas siblings of this node are appended as right successors in the binary representation. The intuition of this encoding is illustrated on Figure~\ref{fig:binary-encoding} for a sample tree.
Trees can be seen as terms or function calls. More formally, a binary tree $t$ can be defined by the recursive syntax $t \ourColoneqq \sigma(t,t') \mid \epsilon $ where $\sigma$ is a node label and $\epsilon$ denotes the empty tree. Similarly unranked trees can be defined as  $t \ourColoneqq \sigma(h) $
where $h$ is a hedge (a sequence of unranked trees) defined as $h \ourColoneqq \sigma(h), h' \mid \epsilon$. The function $f$ that translates unranked trees into binary trees is then defined by $f(\sigma(h), h') = \sigma(f(h), f(h'))$ and $f(\epsilon)=\epsilon$. The reverse mapping used for reconstructing unranked trees from binary trees can be expressed as: $f^{-1}(\sigma(t,t')) = \sigma(f^{-1}(t)), f^{-1}(t')$ and $f^{-1}(\epsilon) = \epsilon$.  

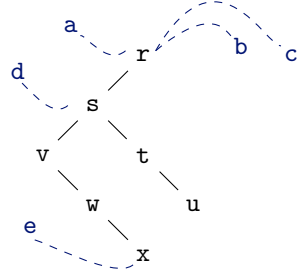
\begin{figure}
\begin{center}
\begin{tikzpicture}[scale=1.1] at (0,0)

\begin{scope}[xshift=0cm, yshift=-4.5cm, scale=0.6]
	
\begin{scope}[xshift=-0.25cm, yshift=0cm]	

	\draw [\attcolor, dashed]  (14.9,3.6) .. controls +(40:-0.5cm) and +(40:0.1cm) .. (14,3.9); 
	\draw (13.8, 4.1) node (att3) [text= \attcolor] {\labelstyle{a}};

	\draw [\attcolor, dashed]  (15.5,3.6) .. controls +(40:1.5cm) and +(-40:-0.1cm) .. (17,3.9); 
	\draw (17.2, 3.7) node (att1) [text= \attcolor] {\labelstyle{b}};
	
	\draw [\attcolor, dashed]  (15.5,3.6) .. controls +(60:2.5cm) and +(4:-0.3cm) .. (18,3.7); 
	\draw (18.2, 3.5) node (att2) [text= \attcolor] {\labelstyle{c}};
	
\end{scope}

	\draw [\attcolor, dashed]  (13.5,2.5) .. controls +(40:-0.5cm) and +(40:0.1cm) .. (12.5,3); 
	\draw (12.5, 3.2) node (att4) [text= \attcolor] {\labelstyle{d}};

\begin{scope}[xshift=1.3cm, yshift=-3.2cm]	
	\draw [\attcolor, dashed]  (13.5,2.5) .. controls +(40:-0.5cm) and +(40:0.1cm) .. (11.5,3); 
	\draw (11.45, 3.25) node (att4) [text= \attcolor] {\labelstyle{e}};
\end{scope}

	\draw (15, 3.5) node (r) [text=black] {\labelstyle{r}};

	\draw (14, 2.5) node (s1) [text=black] {\labelstyle{s}};
	\draw (15, 1.5) node (s2) [text=black] {\labelstyle{t}};		
	\draw (16, 0.5) node (s3) [text=black] {\labelstyle{u}};

	\draw (13, 1.5) node (f1) [text=black] {\labelstyle{v}};
	\draw (14, 0.5) node (f2) [text=black] {\labelstyle{w}};
	\draw (15, -0.5) node (f3) [text=black] {\labelstyle{x}};
	
	\draw (r) -- (s1);
	\draw (s1) -- (s2);
	\draw (s2) -- (s3);
	\draw (s1) -- (f1);
	\draw (f1) -- (f2);
	\draw (f2) -- (f3);	 
	
\end{scope}
\end{tikzpicture}
	\end{center}
\caption{Binary Encoding of Tree of Figure~\ref{fig:xml-tree}.}\label{fig:binary-tree}
\end{figure}

In the remaining part of this manual, the binary representation of a tree is implicitly considered, unless stated otherwise. From an XML point of view, notice that only the nested structure of XML elements (which are ordered) is encoded into binary form like this. XML attributes (which are unordered) are left unchanged by this encoding.  For instance, Figure~\ref{fig:binary-tree} presents how the sample tree of Figure~\ref{fig:xml-tree} is mapped.


\subsection{Syntax of Logical Formulas} \label{formulas-syntax}

\paragraph{Modal Formulas for Navigating in Trees}


The logic uses two  \emph{programs} for navigating in binary trees: the program \texttt{1} allows to navigate from a node down to its first successor and the program \texttt{2} for navigating from a node down to its second successor. The logic also features \emph{converse programs} \texttt{-1} and \texttt{-2} for navigating upward in binary trees, respectively from the first and second successors to the parent node. 
Some basic logical formulas together with corresponding satisfying binary trees are shown on Table~\ref{sample-modal-formulas}.
When using XPath expressions, like e.g. \texttt{select("a[b]")}, the XPath expression is automatically compiled into a logical formula over the binary tree representation (see Section~\ref{supported-xpath}).

\begin{table}
\begin{tabular}{|c|c|c|}
\hline
\textbf{Sample Formula} & \textbf{Satisfying Binary Tree} & \textbf{XML syntax} \\ 
\hline 
\texttt{a \& <1>b}
&
\begin{tikzpicture}[scale=0.75]
\draw (0,1) node (a) {\textbf{a}};
\draw (-1,0) node (b) {b};
\draw [black] (a) -- (b);
\end{tikzpicture}
&
\texttt{<a><b/></a>} \\
\hline
\texttt{a \& <2>b}
&
\begin{tikzpicture}[scale=0.75]
\draw (0,1) node (a) {\textbf{a}};
\draw (1,0) node (b) {b};
\draw [black] (a) -- (b);
\end{tikzpicture}
&
\texttt{<a/><b/>} \\
\hline
\texttt{a \& <1>(b \& <2>c)}
&
\begin{tikzpicture}[scale=0.75]
\draw (0,1) node (a) {\textbf{a}};
\draw (-1,0) node (b) {b};
\draw (0,-1) node (c) {c};
\draw [black] (a) -- (b) -- (c);
\end{tikzpicture}
&
\texttt{<a><b/><c/></a>} \\
\hline
\texttt{e \& <-1>(d \& <2>g)}
&
\begin{tikzpicture}[scale=0.75]
\draw (0,1) node (d) {d};
\draw (-1,0) node (e) {\textbf{e}};
\draw (1,0) node (g) {g};
\draw [black] (g) -- (d) -- (e);
\end{tikzpicture}
&
\texttt{<d><e/></d><g/>} \\
\hline

\texttt{f \& <-2>(g \& \~{}<2>T)}
&
none
&
none \\
\hline
\end{tabular}
\caption{Sample Formulas using Modalities.}\label{sample-modal-formulas}
\end{table}

 


The set of logical formulas is defined by the syntax given on Figure~\ref{fig:logic-syntax}, where the meta-syntax $\oneormoresep{X}$ means one or more occurences of $X$ separated by commas.  Models of a formula are finite binary trees for which the formula is satisfied at some node. The semantics of logical formulas is formally defined in \cite{geneves-phd06,geneves-pldi07}. Table~\ref{sample-modal-formulas} gives basic formulas that use modalities for navigating in binary trees and node names.

\begin{figure}
\begin{center}
\smallsyntax{
\entry      \phi       [formula]
   	      \ctrue [true]	
\oris	      \cfalse [false]
\oris        \nodelabelvar [element name]
\oris         \atomprop [atomic proposition ]
\oris        \ccontextsymb [start context] 
\oris        \phi \corop \phi [disjunction]
\oris        \phi \candop \phi [conjunction]
\oris        \phi \cimplies \phi [implication]
\oris        \phi \cequiv \phi [equivalence]
\oris       \texttt{(} \phi \texttt{)} [parenthesized formula]
\oris        \cneg \phi [negation]
\oris        \cemod{\programvar}\phi  [existential modality]
\oris        \cemod{\nodelabelvar}\ctrue  [attribute named $\nodelabelvar$]
\oris        \cvar [variable]
\oris        \clet ~ \oneormoresep{\cvar= \phi} ~ \cin~ \phi [binder for recursion]
\oris 	      \cpredicatevar [predicate (See Figure~\ref{fig:predicates-syntax})] \\
\entry \programvar [program inside modalities]
\cprog{1} [first child]
\oris \cprog{2} [next sibling]
\oris \cprog{-1} [parent]
\oris \cprog{-2} [previous sibling]
}
\end{center}
\caption{Syntax of Logical Formulas.}\label{fig:logic-syntax}
\end{figure}

\paragraph{Recursive Formulas}
The logic allows expressing recursion in trees through the use of a fixpoint operator.  For example the recursive formula: \begin{center}\texttt{let \$X = b | <2>\$X in \$X}\end{center} means that either the current node is named \texttt{b} or there is a sibling of the current node which is named \texttt{b}.  For this purpose, the variable \texttt{\$X} is bound to the subformula \texttt{b | <2>\$X} which contains an occurence of \texttt{\$X} (therefore defining the recursion). The scope of this binding is the subformula that follows the ``\texttt{in}'' symbol of the formula, that is \texttt{\$X}. The entire formula can thus be seen as a compact recursive notation for a infinitely nested formula of the form: \begin{center}\texttt{b | <2>(b | <2>(b | <2>(...)))}\end{center} 

\noindent Recursion allows expressing global properties. For instance, the recursive formula: \begin{center}\texttt{\~{} let \$X = a | <1>\$X | <2>\$X in \$X} \end{center} expresses the absence of  nodes named \texttt{a} in the whole subtree of the current node (including the current node). Furthermore, the fixpoint operator makes possible to bind several variables at a time, which is specifically useful for expressing mutual recursion. For example, the mutually recursive formula: \begin{center}\texttt{let \$X = (a \& <2>\$Y) | <1>\$X | <2>\$X, \$Y =  b | <2>\$Y in \$X}\end{center} asserts that there is a node somewhere in the subtree such that this node is named \texttt{a} and it has at least one sibling which is named \texttt{b}.  Binding several variables at a time provides a very expressive yet succinct notation for expressing mutually recursive structural patterns (that may occur in DTDs for instance).

The combination of modalities and recursion makes the logic one of the most expressive (yet decidable) logic known. For instance, regular tree grammars can be expressed with the logic using recursion and (forward) modalities. The combination of  converse programs and recursion allows expressing properties about  ancestors of a node for instance. The possibility of nesting recursive formulas allow XPath expressions to be translated into the logic.

\paragraph{Cycle-Freeness Restriction} 
There is a restriction on the use of recursive formulas. Only formulas that are \emph{cycle-free} are allowed.
Intuitively a formula is cycle-free if it does not contain both a program and its converse inside the \emph{same} recursion. For instance, the formula \begin{center}\texttt{let \$X = a | <-1>\$X | <1>\$X in \$X}\end{center}  is not cycle-free since \texttt{1} and \texttt{-1} occur in front of the same variable bound by the same binder. A formula is cycle-free if one cannot find both a program and its converse by starting from a variable and going up in the formula tree to the binder of this variable. For instance, the following formula is cycle-free:
\begin{center}\texttt{let \$X = a \& (let \$X = b | <1>\$X in \$X) | <-1>\$X in \$X}\end{center} since variable binders are properly nested and a program and its converse never appear in front of the same variable bound by the same binder.

Translations of XPath expressions and XML tree types into the logic \emph{always} generate cycle-free formulas, whatever the translated XPath or XML type is. The cycle-freeness restriction only matters when one directly writes recursive logical formulas. From a theoretical perspective the cycle-freeness restriction comes from the fact that converse programs may interact with recursion in a subtle manner such that the finite model property is lost,  so the cycle-freeness restriction ensures that the negation of every formula can also be expressed in the logic, or in other terms, that the logic is closed under negation and all other boolean operations (a detailed discussion on this topic can be found in \cite{geneves-pldi07}).

\begin{figure}
\begin{center}
\smallsyntax{ 
\entry \cpredicatevar [] 
\cselect{\exprvar} [] 
\oris        \cselectcontext{\exprvar}{\phi} []
\oris        \cexists{\exprvar} []
\oris        \cexistscontext{\exprvar}{\phi} [] \\
\oris       \ctype{\filenamevar}{\nodelabelvar} [] 
\oris       \ctypetag{\filenamevar}{\nodelabelvar}{\phi}{\phi'} [] 
\oris       \cforwardincompatible{\phi}{\phi'}   []
\oris       \cbackwardincompatible{\phi}{\phi'}   []\\
\oris       \celem{\phi} []
\oris       \cattr{\phi} [] 
\oris       \cdescendant{\phi} []
\oris       \cexclude{\phi} []
\oris       \caddedelement{\phi}{\phi'} []
\oris       \caddedattribute{\phi}{\phi'} [] \\
\oris       \cnonempty{\exprvar}{\phi} []
\oris       \ccnewelementnames{\exprvar}{\filenamevar}{\filenamevar'}{\nodelabelvar}  []
\oris       \ccnewregions{\exprvar}{\filenamevar}{\filenamevar'}{\nodelabelvar}   []
\oris       \ccnewcontents{\exprvar}{\filenamevar}{\filenamevar'}{\nodelabelvar}  []
\oris \custompredicatevar(\oneormoresep{\phi}) []
}
\end{center}
\caption{Syntax of Predicates for XML Reasoning.}\label{fig:predicates-syntax}
\end{figure}

\begin{figure}
\begin{center}
\smallsyntax{
\entry \logicalspec []
 \phi [ formula (see Fig.~\ref{fig:logic-syntax})]
\oris \predicatedefinitions; \phi []\\
\entry \predicatedefinitions []
 \custompredicatevar(\oneormoresep{\nodelabelvar}) = \phi'  [custom definition]
\oris \predicatedefinitions; \predicatedefinitions [list of definitions]
}
\end{center}
\caption{Global Syntax for Specifying Problems.}\label{fig:lang-syntax}
\end{figure}

\paragraph{Supported XPath Expressions}
 \label{supported-xpath}
The logic provides high-level constructions for facilitating the formulation of problems involving XPath expressions. The construct $\cxpathcontext{e}{\phi}$ where $e$ is an XPath expression provides a way of embedding XPath expression directly into the logic ($e$ is automatically compiled into a logical formula, see  \cite{geneves-pldi07} for details on the compilation technique). The second parameter $\phi$ denotes the context from which the XPath is applied; it can be any formula. The other construct \cxpath{e} is simply a shorthand for $\cxpathcontext{e}{\ccontextsymb}$, where $\ccontextsymb$ is the initial context node mark. The syntax of supported XPath expressions is given on Figure~\ref{fig:xpath-syntax}. We observed that, in practice, many XPath expressions contain syntactic sugars that can also fit into this fragment.
Figure~\ref{fig:xpath-sugars} presents how our XPath parser rewrites some commonly found XPath patterns into the fragment of Figure~\ref{fig:xpath-syntax}, where the notation $\multistep{ \step{\axisvar}{\nodetestvar}}{k}$ stands for the composition of $k$ successive path steps of the same form:
$\underbrace{ \step{\axisvar}{\nodetestvar}/.../ \step{\axisvar}{\nodetestvar}}_{k ~ \text{steps}}$.

\begin{figure}
\smallsyntax{
\entry   \exprvar     []
             /\locpathvar              [absolute path]
\oris        \locpathvar              [relative path]
\oris        \exprvar \mid \exprvar [union] 
\oris        \exprvar \pinter \exprvar [intersection] \\
\entry      \locpathvar          []
             \locpathvar/\locpathvar      [path composition]
\oris        \qualif{\locpathvar}{\qualifvar} [qualified path]
\oris        \step{\axisvar}{\nodetestvar} [step] \\
\entry       \qualifvar         []
    					\qualifvar \op{and} \qualifvar [conjunction]
\oris					\qualifvar \op{or} \qualifvar  [disjunction]
\oris					\xpathfun{not}{\qualifvar}   [negation]
\oris					\locpathvar            [path] 
\oris					\locpathvar/@{\nodetestvar}             [attribute path] 
\oris					@{\nodetestvar}           [attribute step] 
\\
\entry      \nodetestvar          [node test]
            \sigma      [node label]
\oris        * [any node label] \\
\entry    \axisvar [tree navigation axis]
             \axis{self}  
     \mid \axis{child} 
     \mid  \axis{parent}[]
\oris       \axis{descendant} 
\mid        \axis{ancestor} []
\oris  \axis{descendant-or-self} []
\oris  \axis{ancestor-or-self}  []
\oris       \axis{following-sibling} []
\oris  \axis{preceding-sibling}  []
\oris       \axis{following} 
\mid \axis{preceding} [] \\
}
\caption{XPath Expressions.}\label{fig:xpath-syntax}
\end{figure}

\begin{figure}
\begin{align*}
\qualif{\nodetestvar}{\xpathfun{position}{}=1} ~&~ \rewritesinto \qualif{\nodetestvar}{\xpathfun{not}{\step{\axis{preceding-sibling}}{\nodetestvar}}} \\
\qualif{\nodetestvar}{\xpathfun{position}{}=\xpathfun{last}{}} ~&~ \rewritesinto \qualif{\nodetestvar}{\xpathfun{not}{\step{\axis{following-sibling}}{\nodetestvar}}} \\
\qualif{\nodetestvar}{\xpathfun{position}{}=\underbrace{k}_{k>1}}  ~&~ \rewritesinto  \qualif{\nodetestvar}{ \multistep{\step{\axis{preceding-sibling}}{\nodetestvar}}{k-1}} \\
\xpathfun{count}{\locpathvar}=0 ~&~ \rewritesinto \xpathfun{not}{\locpathvar} \\
\xpathfun{count}{\locpathvar}>0 ~&~ \rewritesinto \locpathvar\\
\xpathfun{count}{\nodetestvar}>\underbrace{k}_{k>0} ~&~ \rewritesinto \nodetestvar/\multistep{\step{\axis{following-sibling}}{\nodetestvar}}{k}
\end{align*}
\begin{multline*}
\qualif{\step{\axis{preceding-sibling}}{*}}{\xpathfun{position}{}=\xpathfun{last}{} \op{and} \qualifvar } \\ \rewritesinto \qualif{\step{\axis{preceding-sibling}}{*}}{\xpathfun{not}{\step{\axis{preceding-sibling}}{*}} \op{and} \qualifvar } 
\end{multline*}
\caption{Syntactic Sugars and their Rewritings.}\label{fig:xpath-sugars}
\end{figure}

\paragraph{Supported XML Types}
The logic is expressive enough to allow for the encoding of any regular tree grammar.  The logical construction $\ctype{\emphtext{filename}}{\emphtext{start}}$ provides a convenient way of referring to tree grammars written in usual notations like DTD, XML Schema, or Relax NG. The referred tree type is automatically parsed and compiled into the logic, starting from the given \emphtext{start} symbol (which can be the root symbol or any other symbol defined by the tree type).

\subsection{Predicates}

We build on the aforementioned query and schema compilers, and define additional predicates that facilitate the formulation of decision problems at a higher level of abstraction. Specifically, these predicates are introduced as logical macros with the goal of allowing system usage while focusing (only) on the XML-side properties, and keeping underlying logical issues transparent for the user. Ultimately, we regard the set of basic logical formulas (such as modalities and recursive binders) as an assembly language, to which  predicates are translated. Some built-in predicates include:
\begin{itemize}
\item ....
\item the predicate $\cexclude{\phi}$ which is satisfiable iff there is no node that satisfies $\phi$ in the whole tree. This predicate can be used for excluding specific element names or even nodes selected by a given XPath expression.
\item the predicate $\celem{\treetypevar}$ builds the disjunction of all element names occuring in $\treetypevar$.
\item the predicate $\cdescendant{\phi}$ forces the existence of a node satisfying $\phi$ in the subtree, and $\custompredicatevar(\oneormoresep{\phi})$ is a call to a custom predicate, as explained in the next section.
\end{itemize}

\subsection{Custom Predicates}

Following the spirit of predicates presented in the previous section, users may also define their own custom predicates.  The full syntax of XML logical specifications to be used with the system is defined on Figure~\ref{fig:lang-syntax}, where the meta-syntax $\oneormoresep{X}$ means one or more occurrence of $X$ separated by commas.
A global problem specification can be any formula (as defined on Figure~\ref{fig:logic-syntax}), or a list of custom predicate definitions separated by semicolons and followed by a formula.
A custom predicate may have parameters that are instanciated with actual formulas when the custom predicate is called (as shown on Figure~\ref{fig:predicates-syntax}). A formula bound to a custom predicate may include calls to other predicates, but not to the currently defined predicate (recursive definitions must be made through the let binder shown on Figure~\ref{fig:logic-syntax}).

\section{Overview of the Background Theory} \label{theory-overview}

The logic and its solver are formally described in \cite{geneves-phd06,geneves-pldi07}. The logic is a modal logic of trees, more specifically an alternation-free $\mu$-calculus with converse for finite trees. The logic is equipped with forward and backward modalities, which are notably useful for capturing all XPath (including reverse) axes. The logic is also equipped with a fixed-point operator for expressing recursion in finite trees. A n-ary fixed-point operator is also provided so that mutual recursion occurring in XML types can be succintly expressed in the logic. The logic is also able to express any propositional property, for instance about nodes labels (XML element and attribute names).  Last but not least, the logic is closed under negation \cite{geneves-phd06,geneves-pldi07}, that is, the negation of any logical formula can be expressed in the logic too (this property is essential for checking XPath containment which corresponds to logical implication). All these features together: propositions, forward and backward modalities, recursion (fixed-points operators), and boolean connectives yield a logic of very high expressive power. 
Actually, this logic is one of the most expressive yet decidable known logic. It can express properties of regular tree languages. Specifically, it is as expressive as tree automata (which notably provide the foundation for the Relax NG language in the XML world) and monadic second-order logic of finite trees (often referred as WS2S or ``MSO'' in the literature)  \cite{thatcher68,doner70}. However, the logical solver is considerably (orders of magnitude) faster than solvers for monadic second-order logic, like e.g., the MONA solver \cite{mona-tool} (the MONA solver nevertheless remains useful when one wants to write logical formulas using MSO syntax). Technically, the truth status of a logical formula (satisfiable or unsatisfiable) is automatically determined in exponential time, and more specifically in time $2^{O(n)}$ where $n$ is proportional to (and smaller than) the size of the logical formula \cite{geneves-phd06,geneves-pldi07}. In comparison, the complexity of monadic second-order logic is much higher: it was proved in the late 1960s that the best decision procedure for monadic second order logic is at least hyper-exponential in the size of the formula \cite{thatcher68,doner70} that is, not bounded by any stack of exponentials. The tree logic described in this document currently offers the best balance known between expressivity and complexity for decidability. The acute reader may notice that the complexity of the logic is optimal since it subsumes tree automata and less expressive logics such as CTL \cite{clarke81}, for instance.

XPath expressions and regular tree types can be linearly translated into the logic. This observation allows to generalize the complexity of the algorithm for solving the logic to a wide range of problems in the XML world.

The decision procedure for the logic is based on an inverse tableau method that searches for a satisfying tree. The algorithm has been  proved sound and complete in \cite{geneves-phd06,geneves-pldi07}. The solver is implemented using symbolic techniques like binary decision diagrams (BDDs) \cite{bryant86}. It also uses numerous optimization techniques such as on-the-fly formula normalization and simplification, conjunctive partitioning, early quantification.

Finally, another benefit of this method (illustrated in Section~\ref{example-xpath-sat}) is that the solver can be used to generate an example (or counter-example) XML tree for a given property, which allows for instance to reproduce a program's bug in the developer environment, independently from the logical solver.

\bibliographystyle{named}
\bibliography{references}

\end{document}

%% file: Markup.tex
\renewcommand{\phi}{\varphi}

\newcommand{\ourColoneqq}{::=}

\newcommand{\step}[2]{\text{{#1}::}{#2}}
\newcommand{\axis}[1]{\text{#1}}
\newcommand{\axisvar}{\emph{a}}

\newcommand{\pinter}{\cap}
\newcommand{\qualif}[2]{{#1}\text{[}{#2}\text{]}}
\newcommand{\op}[1]{\mathbin{\text{\small {#1}}}}

\newcommand{\atomprop}{p}

\newcommand{\rewritesinto}{\rightsquigarrow}

\newcommand{\syntaxdef}{\mathrel{::=}}

\newcommand{\syntaxtable}[1]{
  \def\entry##1[##2]##3[##4]{
    {##1} & \syntaxdef & \hspace{3cm} & \!\!\!\! \mbox{##2}
    \\    &     & {##3} & \mbox{##4} }
  \def\singleentry##1[]##2[##3]{
  {##1} & \syntaxdef & {##2} & \!\!\!\! \mbox{##3} }
  \def\oris##1[##2]{
    \\    & |   & {##1} & \mbox{##2} }
  \def\orisopt##1[##2]{
    \\ \left(   & |   & {##1} & \mbox{##2} \right) }
  \begin{array}{rcll}
  #1
  \end{array}
  }
\newcommand{\smallsyntax}[1]{\[\syntaxtable{#1}\]}

\newcommand{\treetypevar}{\mathcal{T}}

\newcommand{\cxpath}[1]{\texttt{select}(\text{"}#1\text{"})}
\newcommand{\cxpathcontext}[2]{\texttt{select}(\text{"}#1\text{"}, #2)}

\newcommand{\ie}{{\text{{\emph{i.e.}}}} }

\newcommand{\nodetestvar}{\text{\emph{nt}}}
\newcommand{\locpathvar}{\emph{path}}

\newcommand{\qualifvar}{\emph{qualifier}}
\newcommand{\exprvar}{\emph{query}}

\newcommand{\xpathfun}[2]{\text{#1}({#2})}

\newcommand{\nodelabelvar}{l}

\newcommand{\programvar}{p}

\newcommand{\ctrue}{\texttt{T}}
\newcommand{\cfalse}{\texttt{F}}

\newcommand{\cimplies}{\text{ }\texttt{=>}\text{ }}
\newcommand{\cequiv}{\text{ }\texttt{<=>}\text{ }}

\newcommand{\cneg}{\tilde{}}

\newcommand{\corop}{\text{ }\texttt{|}\text{ }}
\newcommand{\candop}{\text{ }\texttt{\&}\text{ }}
\newcommand{\ccontextsymb}{\texttt{\#}}

\newcommand{\cprog}[1]{\texttt{#1}}

\newcommand{\cvar}{\texttt{\$X}}

\newcommand{\clet}{\texttt{let}}
\newcommand{\cin}{\texttt{in}}
\newcommand{\emphtext}[1]{\text{\emph{#1}}}
\newcommand{\cemod}[1]{\texttt{<}#1\texttt{>}}

\newcommand{\cpredicatevar}{\text{\emph{predicate}}}

\newcommand{\oneormoresep}[1]{\langle #1 \rangle^\varoplus}

\newcommand{\cselect}[1]{\texttt{select}(\texttt{"}#1\texttt{"})}
\newcommand{\cselectcontext}[2]{\texttt{select}(\texttt{"}#1\texttt{"}, #2)}
\newcommand{\cexists}[1]{\texttt{exists}(\texttt{"}#1\texttt{"})}
\newcommand{\cexistscontext}[2]{\texttt{exists}(\texttt{"}#1\texttt{"}, #2)}

\newcommand{\ctype}[2]{\texttt{type}(\texttt{"}#1\texttt{"}, #2)}
\newcommand{\ctypetag}[4]{\texttt{type}(\texttt{"}#1\texttt{"}, #2, #3, #4)}

\newcommand{\filenamevar}{\text{\emph{f}}}

\newcommand{\cnonempty}[2]{\texttt{non\_empty}(\texttt{"}#1\texttt{"}, #2)}

\newcommand{\celem}[1]{\texttt{element}(#1)}
\newcommand{\cattr}[1]{\texttt{attribute}(#1)}
\newcommand{\caddedelement}[2]{\texttt{added\_element}(#1,#2)}
\newcommand{\caddedattribute}[2]{\texttt{added\_attribute}(#1,#2)}
\newcommand{\cexclude}[1]{\texttt{exclude}(#1)}
\newcommand{\cforwardincompatible}[2]{\texttt{forward\_incompatible}(#1,#2)}
\newcommand{\cbackwardincompatible}[2]{\texttt{backward\_incompatible}(#1,#2)}

\newcommand{\ccnewelementnames}[4]{\texttt{new\_element\_name}(\texttt{"}#1\texttt{"},\texttt{"}#2\texttt{"},\texttt{"}#3\texttt{"}, #4)}
\newcommand{\ccnewregions}[4]{\texttt{new\_region}(\texttt{"}#1\texttt{"},\texttt{"}#2\texttt{"},\texttt{"}#3\texttt{"}, #4)}

\newcommand{\ccnewcontents}[4]{\texttt{new\_content}(\texttt{"}#1\texttt{"},\texttt{"}#2\texttt{"},\texttt{"}#3\texttt{"}, #4)}

\newcommand{\cdescendant}[1]{\texttt{descendant}(#1)}

\newcommand{\custompredicatevar}{\text{\emph{predicate-name}}}

\newcommand{\multistep}[2]{(#1)^{#2}}

\newcommand{\logicalspec}{\text{\emph{spec}}}
\newcommand{\predicatedefinitions}{\text{\emph{def}}}